# Experiences with Content Development and Assessment Design in the Era of GenAI

Aakanksha Sharma, Samar Shailendra, and Rajan Kadel
*SITE, Melbourne Institute of Technology, Australia*

*Abstract*—Generative Artificial Intelligence (GenAI) has the potential to transform higher education by generating human-like content. The advancement in GenAI has revolutionised several aspects of education, especially subject and assessment design. In this era, it is crucial to design assessments that challenge students and cannot be solved using GenAI tools. This makes it necessary to update the educational content with rapidly evolving technology. The assessment plays a significant role in ensuring the students' learning, as it encourages students to engage actively, leading to the achievement of learning outcomes. The paper intends to determine how effectively GenAI can design a subject, including lectures, labs and assessments, using prompts and custom-based training. This paper aims to elucidate the direction to educators so they can leverage GenAI to create subject content. Additionally, we provided our experiential learning for educators to develop content, highlighting the importance of prompts and fine-tuning to ensure output quality. It has also been observed that expert evaluation is essential for assessing the quality of GenAI-generated materials throughout the content generation process.

## I. INTRODUCTION

The traditional approach to education often adopts uniform teaching methodologies, failing to address the individual diverse and unique learning needs. This approach adheres to a "one size fits all" ideology, neglecting the individual's learning style and capabilities. These limitations can be overcome with GenAI. It assists in teaching, research, writing, and personalised learning in education for improving student learning [1], [2]. Furthermore, GenAI uses machine learning algorithms to address queries by responding to the prompts and generating content in various formats, including audio, text, images, code, and videos. It can create custom assessments (individual or group-based), leading to better engagement and effective learning. Thus, educators must design/redesign the teaching content using GenAI to accommodate the student's individual needs, and it will enable students to experience the potential of GenAI in their learning. Currently, some educators are using it, while others are still concerned with the uncertainties associated with using GenAI, such as academic integrity and achieving student learning outcomes [3]. Despite all the opportunities, this technology also has some challenges, such as over-reliance on technology [4], preserving academic integrity [5]–[7], cognitive bias [6], privacy [8], [9] and lack of regulation [10]. Moreover, GenAI is not universally accessible, thus leading to accessibility challenges [4], [6]. The training of GenAI models poses a significant challenge.

These models are trained on vast data and if the data is noisy, has errors and/or contains biases then the content generated by GenAI will not be factual or be incorrect. This is often termed as hallucination [11], [12]. Due to these challenges, an expert instructor and a thorough evaluation of the content for correctness are necessary throughout the content generation process. Overall, GenAI brings legitimate opportunities to use it. Thus, the United Nations Educational, Scientific, and Cultural Organisation (UNESCO) recommends using GenAI in education to utilise its potential benefits [13], [14].

### A. GenAI and Prompt Engineering

GenAI models are built on cutting-edge technologies, such as Transformers [15], which has advanced natural language processing. GenAI applications use different Large Language Models (LLMs) such as GPT, LLama etc. to generate responses for the user prompts. ChatGPT [16], one of the most talked about GenAI tool, is based on GPT. These LLMs need to be trained on massive data and their responses are also based on the data being used to train these models. Most of the current LLMs are based on transformer architecture. The transformers use the self-attention mechanism to process and generate the language better than earlier models. These transformer models can be classified as encoder based e.g. BERT [17] and CodeBERT [18], a decoder based e.g. GPT [19] and Llama [20] or encoder-decoder based e.g. BART [21] and CodeT5 [22], depending upon the transformer architecture. GPT-4 can be used for diverse tasks such as writing, coding and creating custom content based on prompts.

In this study, Custom Data Fine-Tuning and Prompt-based Fine-Tuning are used to generate the subject content. By combining these approaches, we can potentially achieve better results with less training data and reduced time. These approaches are discussed in Section III of the paper.

### B. Motivation and Contribution

The key motivation behind this study is to understand that traditional teaching methodology is not catered to meet the individual needs of students, while with GenAI, educators can leverage and generate tailored content based on their students' needs and capabilities. We have considered the diversity of subjects and experimented to generate the lecture, lab, and assessment for different subjects. The important learning is that GenAI can effectively generate the content if model training and prompts are good. We faced some challenges while experimenting with GenAI for content development, including writing bad prompts and training with data having lots of images. In Section IV of the paper, we share our experiences with prompt engineering and provide examples of both good and bad prompts.

The key contributions of this paper can be summarised as:
- Introducing the lecture design process for adopting GenAI, considering the importance of prompts and custom training.
- Introducing the lab design process to effectively generate tailored lab content using GenAI.
- Introducing the assessment design process by ensuring assessment authenticity.

## II. RELATED WORKS

Models such as GPT-4 are trained for general text and content and are good for interactions and general queries. They lose the rigour for special purpose scenarios and only achieve limited precision [23], [23]. These models can be trained for contextual data and fine-tuned for the specific domain [24]–[26]. Fine-tuning a model involves updating the pre-trained model's parameters using unsupervised learning to optimise its output for a specific domain or task. In [27], [28] authors have suggested fine-tuning LLMs for the manufacturing domain.

The methods of content development, assessments and lab designs in the GenAI era have been extensively discussed in the literature [29], [30]. The authors in [31], [32] have discussed the impact on assessment design and innovation required for instructional Design in the GenAI era. However, these models can hallucinate and generate unrelated and irrelevant information which may not be relevant to the context. The researchers have suggested to exercise caution while using these models [33], [34].

From the literature, it can be understood that the pre-trained general-purpose LLMs are not optimal for custom purposes. Fine-tuning for the education-specific requirements such as content development, lab and assessment design is a desirable approach. Hence there is a burgeoning need to develop a methodology and framework to fine-tune these LLMs for custom educational purposes using subject matter content.

## III. DESIGN APPROACH

Large Language Models can engage with us in natural language across various fields that are most familiar to us. Interestingly, these LLMs can be made to impersonate any personality using prompts to respond to a particular situation (e.g. they can be prompted to act as a student or a teacher) and also be fine-tuned using custom data for specific scenarios or use cases.

These LLMs like GPT-4, LLama, etc. are trained over large datasets and further provide the capability to be fine-tuned to suit specific purposes. The performance of these models can be improved using two approaches: i) Custom Data Fine-Tuning ii) Prompt-based Fine-Tuning

***Custom Data Fine-Tuning*** allows us to use the custom data to further fine-tune the model [16]. This ensures that the responses generated by the model are predominantly based on the information available in the custom data. We can use the APIs provided by the LLM provider to fine-tune the model. On the other hand, the ***Prompt-based Fine-Tuning*** has two stages. In the first stage, the model is made to assume certain roles based upon the prompts. In another stage, the model is prompted to generate a response in line with the instructions given. Since the LLMs just complete the sentences with the most probable words, the language and instructions given to the model during the two prompting stages are quite different. The former is designed to make LLM use certain roles and behave in a certain way, while the later prompts are more instructional and oriented for the task to be completed.

In this paper, we have used a combination of these strategies to fine-tune our model for the particular task. We first use the custom dataset relevant to the purpose to fine-tune the model. Furthermore, we use descriptive prompts for the model to impersonate a particular role or behaviour, followed by the prompts to respond to certain tasks. Some examples of sample data and prompts are provided in Section IV.

### A. Lecture Content Design

The process of generating the lecture content using GenAI is discussed in this section. Figure 1 represents the step-by-step process that can be followed to create lectures considering the course description and Unit Learning Outcomes (ULOs). It starts by fine-tuning the model, providing custom-reference material and course details, and prompting the model to suggest subject outline topics, sub-topics and lecture slides. The subject expert plays an essential role in the whole process of generating content. An effective evaluation by experts is required to validate the correctness of the generated content throughout the process. Different steps involved in the lecture design are mentioned below:

1) **Custom Model Tuning:** In this step, the model is fine-tuned using the custom reference material for the course. This ensures the model predominantly uses our custom reference material to generate the content.
   - *Input:* Course Description and Reference Material.
   - *Outcome:* Custom Fine-Tuned model.

2) **Prompting Model for lecture design:** Using the model created in the first step, prompting the model to develop subject outline topics using the key inputs: Course title and description, including the ULOs. We will prompt the model to generate an outline of 10 subject topics.
   a) Mapping of Topic Outline and ULO: In this step, the expert will analyse and evaluate the suggested topics if they match the ULO(s) of the subject.
      - *Input:* Expert Analysis/Evaluation.
      - *Outcome:* Either accept the topics or refine the prompts to generate new topic outlines.
   b) Generate Sub-topics/Lecture slides: If the suggested topics from the previous step are as per the requirements, then generate the sub-topics/Lecture slides. Refine the prompts until the expert accepts the sub-topics, and then generate the lecture slides. Prompts should be refined to get the quality output.
- *Input:* User Prompts, Expert Inputs.
- *Outcome:* Sub-topics and Lecture Slides.

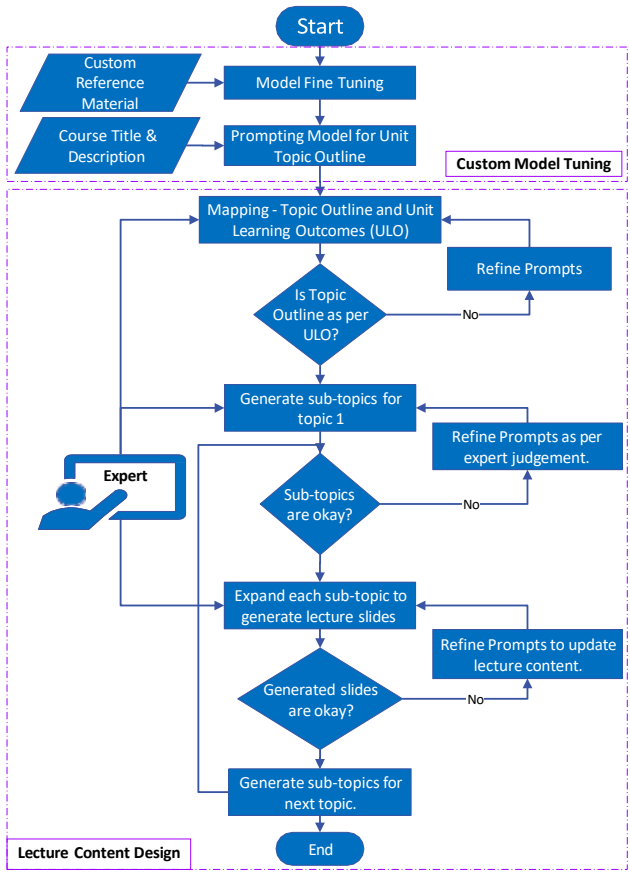

Fig. 1. Lecture content creation process.

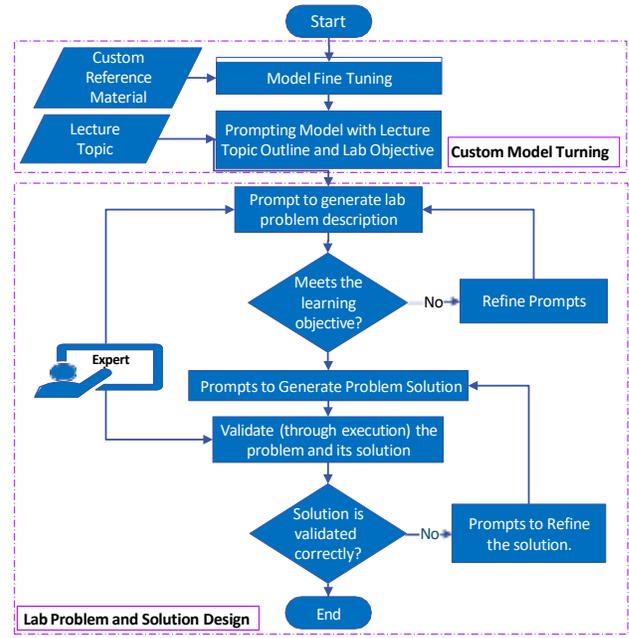

Fig. 2. Lab design process.

Repeat the same process to generate sub-topics and lecture slides for the next topic. The whole process ensures that GenAI helps but currently does not omit the need for expert instructors and thorough content evaluation for correctness.

### B. Lab Design

Figure 2 presents the process that can be followed to generate lab tasks to augment the lecture topic. Different stages involved in lab design are explained below:

1) **Custom Model Tuning:** As described in the previous section (Section III), this step uses the same reference material as being used to train the lecture content to maximise the correlation between the lectures and the labs. We further prepare the model using specific prompts corresponding to the lab topic. This will set the context of the model for the required task. Since the lab work must be aligned with the lecture, we prompt the model with the corresponding lecture topic along with other related instructions.
   - *Input:* Reference material, Lecture Topic.
   - *Outcome:* Custom Fine-Tuned model.

2) **Lab Problem and Solution Design:** At this step, we further prompt the model to generate the lab problem and also generate the solution for the problem. The expert should validate that the solution generated is getting executed correctly on the real system as well.
   - *Input:* User Prompts, Expert Inputs, Validation.
   - *Outcome:* Lab Problem Description and Solution.

### C. Assignment Design

Figure 3 illustrates the assessment creation process utilising subject resources, laboratory resources, learning outcomes specific to the assignment, qualification level and type of assignment. The created assignment needs to be examined by the expert in the field in terms of meeting the accuracy of the content, learning outcomes and qualification level. If any of those requirements are not met, a refine prompt is used to update the created assignment but not a new assignment.

1) **Custom Model Tuning**
   a) Model Fine-Tuning: In this step, the model fine-tuning is completed using reference materials, lecture and laboratory resources as discussed in Section III. At the end of this step, the trained model becomes available for further processing.
      - *Input:* Reference materials, lecture and laboratory resources.
      - *Outcome:* Custom Trained model.

2) **Assessment Design and Validation**
   a) Prompting Model: Using the model created in the first step, prompt for assignment creation is

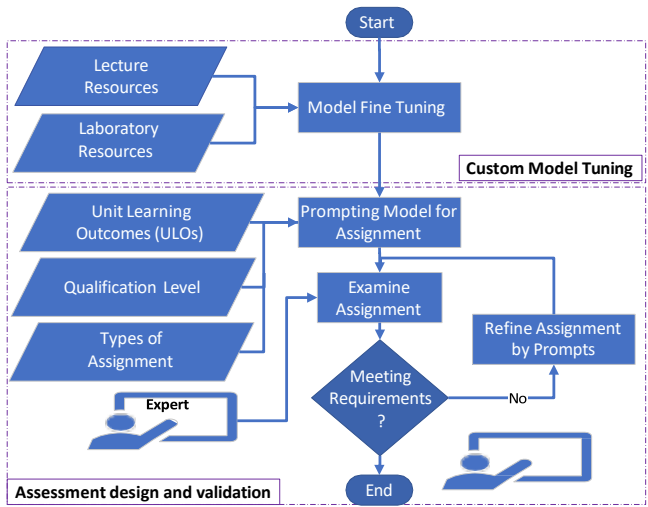

Fig. 3. Assignment creation process.

performed using the key inputs: ULO(s) specific to the assignment, qualification level and type of assessment. Action verbs from Bloom's Taxonomy [35] can be used to ensure the qualification level and learning outcomes.
- *Input:* ULOs specific to assignment, qualification level, Bloom's Taxonomy, type/sample of assignment, and prompt(s)
- *Outcome:* Assignment.

b) Examine Assignment: The assignment created in the previous step will be evaluated using an expert. Based on the expert's judgement, the assignment will either be accepted or require further refinement.
- *Input:* Expert analysis / evaluation.
- *Outcome:* Either accept the assignment or go for assignment refinement.

c) Refine Assignment: In this step, the suitable prompt(s) are used to update the previous version of the assignment considering expert judgement.
- *Input:* Previous version of assignment, expert judgement and specific prompt(s).
- *Outcome:* Refined assignment.

Steps b) and c) continue until the expert is satisfied that the assignment meets the requirements about learning outcomes and qualification level.

## IV. EXPERIENTIAL SHARING

This section shares our experience using GenAI for lecture, lab and assessment design. We aim to provide insights into practical aspects, including key challenges and solutions, that we faced during the subject development process. We designed the Subjects - "Network Automation" and "Enterprise Cloud Networks", using our custom reference material, considering 20 different pdf files (a few e-books, and articles). The complete step-by-step process for lecture, lab and assessment design is discussed in the previous sections. We began by identifying the course's key topics and outlining the subject's structure. With GenAI, we generated detailed lecture slides, notes, lab exercises and assignments using our custom reference material, which is unique for lectures, labs and assignments. The process involved providing specific prompts and iterating on the responses to refine the content. Every time, the generated content is verified by an expert to ensure that the content is aligned with our learning objectives and standards. Key takeaways from our experience include the following:

1) **Iterative Refinement:** GenAI provides many ideas and detailed explanations but requires significant refinement to get the required content.
2) **Content Accuracy:** Ensuring the content's accuracy and relevance is a significant challenge, as the model is trained on custom data, and if the data is inaccurate, then the generated content will most likely be inaccurate, so ensure that the training reference material is accurate.
3) **Content Pre-Processing:** If training reference material has many images, sometimes training a model becomes challenging. In this case, pre-processing training data is a good option.
4) **Role of Prompts:** The content generated by GenAI varies according to the prompts used. Good prompts are essential to generate the required content; otherwise, refine the prompts until you get the desired content. The prompts play a significant role in the content generation process; below are examples of good and bad prompts to understand their importance.

> **Prompt 1:** Example
>
> *Role Assumption Prompt: You are an expert in "Network Automation". I am planning to develop a unit on "Network Automation" to teach Master level students. I will provide you the required ULOs, and reference material. Can you suggest 10 topics to teach in this unit?*

> **Prompt 2:** Example
>
> *Bad Prompt: Can you give me 20 slides for the topic "Network Automation" that I could use to teach my students?*
> *Good Prompt: Please give me 2-3 "Network Automation" slides. The slides should include 4-5 bullet points about the topic and provide a use case or an example. Also, provide me with slide notes to talk about each slide.*

> **Prompt 3:** Example
>
> ***Bad Prompt:*** *Please add more details to slides?*
> ***Good Prompt:*** *Expand on each of the subtopics you provided earlier. You can consider elaborating on the key ideas, offering supporting examples and explaining any details that you think would enhance the audience's understanding of the topic. Also, generate speaker notes. By this, we should have nearly 20 slides on this topic.*

The content generated with good prompts is more detailed and closely aligned with the topic, incorporating examples and lecture notes for teaching, while the content generated using bad prompts is incomplete and inadequate for unit development.

5) **Expert Oversight:** It is important to understand that GenAI should only be considered a tool that can design human-like text, but it is not always right, and its response should be critically evaluated. Therefore, it requires increased capabilities for university teachers to make content judgements.

## V. Conclusions

It is necessary to understand the importance of iterative refinement while working with GenAI and the value of combining AI-generated content with expert review. For educators considering GenAI tools, we recommend starting with clear objectives and being prepared for an iterative process. Engage critically with the GenAI's output and leverage its strengths to enhance, rather than replacing traditional teaching methods. Lastly, every content and assessment generated by GenAI must be thoroughly reviewed by the experts in the respective area before being shared with the students.

## References

[1] A. Bahrini *et al.*, "ChatGPT: Applications, opportunities, and threats," in *2023 Systems and Information Engineering Design Symposium (SIEDS)*. IEEE, 2023, pp. 274–279.
[2] E. Kasneci *et al.*, "ChatGPT for good? On opportunities and challenges of large language models for education," *Learning and individual differences*, vol. 103, p. 102274, 2023.
[3] W. Holmes *et al.*, "Ethics of AI in education: Towards a community-wide framework," *International Journal of Artificial Intelligence in Education*, pp. 1–23, 2021.
[4] P. P. Ray, "ChatGPT: A comprehensive review on background, applications, key challenges, bias, ethics, limitations and future scope," *Internet of Things and Cyber-Physical Systems*, 2023.
[5] G. M. Currie, "Academic integrity and artificial intelligence: is ChatGPT hype, hero or heresy?" *Seminars in Nuclear Medicine*, vol. 53, no. 5, pp. 719–730, 2023, preclinical.
[6] E. Sabzalieva and A. Valentini, "ChatGPT and artificial intelligence in higher education: quick start guide," *International Institute for Higher Education in Latin America and the Caribbean (IESALC)*, 2023.
[7] D. R. Cotton *et al.*, "Chatting and cheating: Ensuring academic integrity in the era of ChatGPT," *Innovations in Education and Teaching International*, pp. 1–12, 2023.
[8] H. Montenegro *et al.*, "Privacy-Preserving Generative Adversarial Network for Case-Based Explainability in Medical Image Analysis," *IEEE Access*, vol. 9, pp. 148 037–148 047, 2021.
[9] M. Gupta *et al.*, "From ChatGPT to ThreatGPT: Impact of Generative AI in Cybersecurity and Privacy," *IEEE Access*, vol. 11, pp. 80 218–80 245, 2023.
[10] C. K. Y. Chan, "A comprehensive AI policy education framework for university teaching and learning," *International Journal of Educational Technology in Higher Education*, vol. 20, no. 1, pp. 1–25, 2023.
[11] Z. Ji *et al.*, "Survey of Hallucination in Natural Language Generation," *Association for Computing Machinery*, vol. 55, no. 12, Mar 2023.
[12] L. Shen *et al.*, "Identifying Untrustworthy Samples: Data Filtering for Open-domain Dialogues with Bayesian Optimization," in *Proceedings of the 30th ACM International Conference on Information & Knowledge Management*, New York, NY, USA, 2021, p. 1598–1608.
[13] W. Holmes *et al.*, *Guidance for generative AI in education and research*. UNESCO Publishing, 2023.
[14] S. Shailendra, R. Kadel, and A. Sharma, "Framework for adoption of generative artificial intelligence (genai) in education," *IEEE Transactions on Education (Early Access)*, pp. 1–9, 2024. [Online]. Available: https://doi.org/10.1109/TE.2024.3432101
[15] A. Vaswani, N. Shazeer *et al.*, "Attention is all you need," *Advances in neural information processing systems*, vol. 30, 2017.
[16] K. I. Roumeliotis and N. D. Tselikas, "Chatgpt and open-ai models: A preliminary review," *Future Internet*, vol. 15, no. 6, p. 192, 2023.
[17] J. Devlin *et al.*, "Bert: Pre-training of deep bidirectional transformers for language understanding," *arXiv preprint arXiv:1810.04805*, 2018.
[18] Z. Feng *et al.*, "Codebert: A pre-trained model for programming and natural languages," *arXiv preprint arXiv:2002.08155*, 2020.
[19] A. Radford *et al.*, "Improving language understanding by generative pre-training," 2018.
[20] H. Touvron *et al.*, "Llama 2: Open foundation and fine-tuned chat models," *arXiv preprint arXiv:2307.09288*, 2023.
[21] M. Lewis *et al.*, "Bart: Denoising sequence-to-sequence pre-training for natural language generation, translation, and comprehension," *arXiv preprint arXiv:1910.13461*, 2019.
[22] Y. Wang *et al.*, "Codet5: Identifier-aware unified pre-trained encoder-decoder models for code understanding and generation," *arXiv preprint arXiv:2109.00859*, 2021.
[23] Y. Sun *et al.*, "LLM4Vuln: A Unified Evaluation Framework for Decoupling and Enhancing LLMs' Vulnerability Reasoning," *arXiv preprint arXiv:2401.16185*, 2024.
[24] W. Ma *et al.*, "Combining Fine-Tuning and LLM-based Agents for Intuitive Smart Contract Auditing with Justifications," *arXiv preprint arXiv:2403.16073*, 2024.
[25] C. Jeong, "Fine-tuning and utilization methods of domain-specific llms," *arXiv preprint arXiv:2401.02981*, 2024.
[26] ——, "Domain-specialized LLM: Financial fine-tuning and utilization method using Mistral 7B," *Journal of Intelligence and Information Systems*, vol. 30, no. 1, p. 93–120, Mar. 2024.
[27] S. Kernan Freire *et al.*, "Harnessing large language models for cognitive assistants in factories," in *Proceedings of the 5th International Conference on Conversational User Interfaces*, 2023, pp. 1–6.
[28] L. Makatura *et al.*, "How Can Large Language Models Help Humans in Design and Manufacturing?" *arXiv preprint arXiv:2307.14377*, 2023.
[29] J. Mao, B. Chen, and J. C. Liu, "Generative Artificial Intelligence in Education and Its Implications for Assessment," *TechTrends*, vol. 68, no. 1, pp. 58–66, 2024.
[30] M. Bower *et al.*, "How should we change teaching and assessment in response to increasingly powerful generative Artificial Intelligence? Outcomes of the ChatGPT teacher survey," *Education and Information Technologies*, pp. 1–37, 2024.
[31] C. B. Hodges and P. A. Kirschner, "Innovation of Instructional Design and Assessment in the Age of Generative Artificial Intelligence," *TechTrends*, pp. 1–5, 2024.
[32] B. K. M. Rajan Kadel *et al.*, "Crafting Tomorrow's Evaluations: Assessment Design Strategies in the Era of Generative AI," *10th International Symposium on Educational Technology (ISET)*, 2024.
[33] Z. Han *et al.*, "An explorative assessment of ChatGPT as an aid in medical education: use it with caution," *Medical Teacher*, vol. 46, no. 5, pp. 657–664, 2024.
[34] M. Sallam, "ChatGPT utility in healthcare education, research, and practice: systematic review on the promising perspectives and valid concerns," in *Healthcare*, vol. 11, no. 6. MDPI, 2023, p. 887.
[35] J. Orrell, "Designing an assessment rubric," *TEQSA*. [Online]. Available: https://www.teqsa.gov.au/sites/default/files/2022-10/designing-assessment-rubric.pdf